\documentclass[useAMS,usenatbib, usegraphicx, fleqn]{mn2e}

\usepackage{graphicx}
\usepackage{epsfig}
\usepackage{hyperref}
\usepackage{amsmath} 
\usepackage{multirow}
\usepackage{amsfonts}
\usepackage{amssymb}
\usepackage{amsxtra,eufrak}
\usepackage{times}
\usepackage[utf8]{inputenc}
\usepackage{pslatex}
\usepackage{color}

\title[ACF large-scale analysis with LGs]{Large-scale analysis of the SDSS-III DR8 photometric luminous galaxies angular correlation function}

\author[de Simoni et al.]
{\parbox{\textwidth}{Fernando de Simoni$^{1,2}$\thanks{Email: fsimoni@id.uff.br},
Fl\'avia Sobreira$^{2,3}$,
Aurelio Carnero$^{2,3}$,
Ashley J. Ross$^{4}$,
Hugo O. Camacho$^{2,5}$,
Rogerio Rosenfeld$^{2,6}$,
Marcos Lima$^{2,5}$,
Luiz A. N. da Costa$^{2,3}$,
Marcio A. G. Maia$^{2,3}$} \vspace{0.4cm}\\
\parbox{\textwidth}{$^1$ Departamento de F\'isica e Matem\'atica, PURO/Universidade Federal Fluminense, Rua Recife s/n, Rio das Ostras, RJ 28890-000, Brazil \\
$^2$ Laborat\'orio Interinstitucional de e-Astronomia - LIneA, Rua Gal. Jos\'e Cristino 77, Rio de Janeiro, RJ 20921-400, Brazil \\
$^3$ Observat\'orio Nacional, Rua Gal. Jos\'e Cristino 77, Rio de Janeiro, RJ 20921-400, Brazil \\
$^4$ Institute of Cosmology and Gravitation, Dennis Sciama Building, University of Portsmouth, Portsmouth PO1 3FX, UK \\
$^5$ Departamento de F\'isica Matem\'atica, Instituto de F\'isica, Universidade de S\~ao Paulo, SP CP 66318, CEP 05314-970,, Brazil \\
$^6$ Instituto de F\'isica Te\'orica and ICTP South American Institute for Fundamental Research, Universidade Estadual Paulista, Rua Dr. Bento T. Ferraz 271, S\~ao Paulo, SP 01140-070, Brazil \\
}}

\begin{document}   
%
%
%
%


\def\aj{\rm{AJ}}                   
\def\araa{\rm{ARA\&A}}             
\def\apj{\rm{ApJ}}                 
\def\apjl{\rm{ApJ}}                
\def\apjs{\rm{ApJS}}               
\def\ao{\rm{Appl.~Opt.}}           
\def\apss{\rm{Ap\&SS}}             
\def\aap{\rm{A\&A}}                
\def\aapr{\rm{A\&A~Rev.}}          
\def\aaps{\rm{A\&AS}}              
\def\azh{\rm{AZh}}                 
\def\baas{\rm{BAAS}}               
\def\jrasc{\rm{JRASC}}             
\def\memras{\rm{MmRAS}}            
\def\mnras{\rm{MNRAS}}             
\def\pra{\rm{Phys.~Rev.~A}}        
\def\prb{\rm{Phys.~Rev.~B}}        
\def\prc{\rm{Phys.~Rev.~C}}        
\def\prd{\rm{Phys.~Rev.~D}}        
\def\pre{\rm{Phys.~Rev.~E}}        
\def\prl{\rm{Phys.~Rev.~Lett.}}    
\def\pasp{\rm{PASP}}               
\def\pasj{\rm{PASJ}}               
\def\qjras{\rm{QJRAS}}             
\def\skytel{\rm{S\&T}}             
\def\solphys{\rm{Sol.~Phys.}}      
\def\sovast{\rm{Soviet~Ast.}}      
\def\ssr{\rm{Space~Sci.~Rev.}}     
\def\zap{\rm{ZAp}}                 
\def\nat{\rm{Nature}}              
\def\iaucirc{\rm{IAU~Circ.}}       
\def\aplett{\rm{Astrophys.~Lett.}} 
\def\apspr{\rm{Astrophys.~Space~Phys.~Res.}}
\def\bain{\rm{Bull.~Astron.~Inst.~Netherlands}} 
\def\fcp{\rm{Fund.~Cosmic~Phys.}}  
\def\gca{\rm{Geochim.~Cosmochim.~Acta}}   
\def\grl{\rm{Geophys.~Res.~Lett.}} 
\def\jcp{\rm{J.~Chem.~Phys.}}      
\def\jgr{\rm{J.~Geophys.~Res.}}    
\def\jqsrt{\rm{J.~Quant.~Spec.~Radiat.~Transf.}}
\def\memsai{\rm{Mem.~Soc.~Astron.~Italiana}}
\def\nphysa{\rm{Nucl.~Phys.~A}}   
\def\physrep{\rm{Phys.~Rep.}}   
\def\physscr{\rm{Phys.~Scr}}   
\def\planss{\rm{Planet.~Space~Sci.}}   
\def\procspie{\rm{Proc.~SPIE}}   

\let\astap=\aap
\let\apjlett=\apjl
\let\apjsupp=\apjs
\let\applopt=\ao


\pagerange{\pageref{firstpage}--\pageref{lastpage}} \pubyear{2011}

\maketitle
\label{firstpage}

\begin{abstract}

We analyse the large-scale angular correlation function (ACF) of the CMASS luminous galaxies (LGs), a photometric-redshift catalogue based on the Data Release 8 (DR8) of the Sloan Digital Sky Survey-III. This catalogue contains over $600 \, \, 000$ LGs in the range $0.45 \leq z \leq 0.65$, which was 
split into four redshift shells of constant width. First, we estimate the constraints on the redshift-space distortion (RSD) parameters $b\sigma_8$ 
and $f\sigma_8$, where $b$ is the galaxy bias, $f$ the growth rate and $\sigma_8$ is the normalization of the perturbations,
finding that they vary appreciably among different redshift shells, in agreement with previous results using DR7 data. 
When assuming constant RSD parameters over 
the survey redshift range, we obtain $f\sigma_8 = 0.69 \pm 0.21$, which agrees at the $1.5\sigma$ level with Baryon Oscillation Spectroscopic Survey DR9 spectroscopic results. 
Next, we performed two cosmological analyses, where relevant parameters not fitted were kept fixed
at their fiducial values. In the first analysis, we extracted the baryon acoustic oscillation peak position for the four redshift shells, 
and combined with the sound horizon scale from 7-year \textit{Wilkinson Microwave Anisotropy Probe} $(WMAP7)$ to produce the constraints $\Omega_{m}=0.249 \pm 0.031$ and 
 $w=-0.885 \pm 0.145$.  In the second analysis, we used the ACF full shape information to constrain cosmology using real data
 for the first time, finding $\Omega_{m} = 0.280 \pm 0.022$ and $f_b = \Omega_b/\Omega_m = 0.211 \pm 0.026$. 
 These results are in good agreement with $WMAP7$ findings, showing that the ACF can be efficiently applied to constrain cosmology 
 in future photometric galaxy surveys.

\end{abstract}


\maketitle

\section{INTRODUCTION}\label{sec:intro}
The study of the large-scale structure of the Universe represents an important
cosmological tool and recent galaxy surveys have become sufficiently large to
competitively constrain cosmological parameters. 
For instance, spectroscopy surveys such as the Two-degree Field Galaxy Redshift Survey \citep[2dFGRS;][]{2dFref} and the Sloan Digital Sky Survey \citep[SDSS;][]{sdssref}
used the 3D galaxy clustering analysis to constrain cosmological parameters. Last year the WiggleZ \citep{wigglez12} released its final cosmological results from galaxy distribution, measuring redshifts out to $z \sim 1$. The Baryon Oscillation Spectroscopic Survey \citep[BOSS,][]{BOSS}, part of the SDSS-III, 
is an ongoing project that is pushing the analysis of the galaxy distribution to another level. It is going to measure more galaxies compared to previous spectroscopic surveys at an effective redshift of $z \sim 0.57$.
Some few representative papers are 
\citet{Percival:2007yw,Blake:2011en,Anderson:2013oza}.

Some of the next generation galaxy surveys will be carried out with photometric data instead of spectroscopy, using 
broad-band photometry to estimate
the so-called photometric redshifts, or photo-zs for short. 
These surveys will estimate photo-zs for a large number of objects, 
but with a lower accuracy compared to spectroscopic redshifts, effectively trading accuracy for statistical power. 
Obviously, this is only possible with a careful characterization of the photo-z errors.  
The typical approach is to slice the survey into redshift shells with thickness of order of the photo-z errors, and study 
the angular clustering on each shell. The 3D information can then be partially restored by also including the 
correlations between different redshift shells and the photo-z errors.

One of the next large photometric surveys is the 
Dark Energy Survey \citep[DES,][]{DESwhite}, which had its first light in 2012 September. 
This survey expects to measure over $\sim 300$ million galaxies within an area of 5000~deg$^2$ of the southern sky up to redshift $z \sim 1.4$. 
Another proposed photometric galaxy survey is the Large Synoptic Survey Telescope (LSST)  with expected science data for 2021 \citep{LSST}. 
This survey will detect over a billion galaxies and will go deeper than the DES in redshift.

Photometric galaxy surveys will demand a full understanding of the angular clustering of the galaxy distribution in order to provide useful cosmological information. 
Therefore several studies have been performed in order to gauge the 
use of the galaxy angular clustering at large scales, 
both on theoretical and observational grounds. 
We briefly review some of them below.

On the theoretical front, 
\citet{simpson09} performed the first study on the measurement of the baryon acoustic oscillation (BAO) peak in the galaxy angular correlation function (ACF) in configuration space using photometric redshifts. 
They emphasized the role of  photo-z errors 
in establishing the connection between the observed BAO position and the sound horizon scale.
\citet{sobreira11} forecasted the cosmological constraints in a DES like survey from the ACF full shape information using the Fisher matrix formalism. 
They found that DES will constrain the dark energy equation of state $w$ with a precision of $\sim 20\%$.
\citet{CrocceEtAl} verified the accuracy of the ACF theoretical covariance matrix against $N-body$ simulations, showing that at scales larger than $\sim 20 \, h^{-1}$Mpc, 
the Gaussian covariance is a good approximation.
\citet{ross11} forecasted constraints on redshift-space distortion (RSD) parameters for a DES like survey from the ACF full shape information and
\citet{Sanchez11} developed a method to apply the BAO peak position in the ACF as a standard ruler, overcoming some issues outlined in \citet{simpson09}.

On the observational front, only one galaxy survey had the characteristics to make it possible to look into the large scale properties of the ACF using photo-zs: the SDSS. This survey produced a series of data releases with four of them leading to a cosmological analysis with photometric data: Data Release 3 \citep[DR3][]{DR3}; the DR4 which was used to produce the MegaZ photometric catalogue \citep{CollisterEtAl}, the DR7  \citep{DR7ref} and the recent DR8 luminous galaxies (LGs) catalogue  \citep{RossEtAl}. These four photometric catalogues resulted in a series of results on the 
angular clustering of galaxies at large scales, 
mostly in the redshift range $0.45 \leq z \leq 0.65$.

\citet{PaddyEtAl} estimated the angular power spectrum in eight redshift shells, constraining RSD parameters and $\Omega_{m}$. 
\citet{Blake07} used the MegaZ catalogue to produce the first cosmological constraints directly from the galaxy angular clustering using the angular power spectrum. 
\citet{Sawangwit11} measured the large-scale ACF but did not constrain cosmological parameters due to an excess power at these scales. \citet{Thomas11a} produced a similar analysis as that of \citet{Blake07}, but for the improved DR7. 
\citet{crocceRSD} used DR7 data to constrain the so-called RSD parameters with 
the ACF full shape information, but did not estimate the cosmology. 
\citet{Carnero12} used the BAO peak position information in DR7 to find the sound horizon scale. 
\citet{RossEtAl} measured the large scale ACF in DR8 in order to check the impact of systematics, 
reducing the excess of power at these scales reported earlier \citep{Sawangwit11,Thomas11b}. 
Using the DR8, \citet{Ho12} estimated the cosmological parameters from the full information of the angular 
power spectrum and 
\citet{Seo12} found the sound horizon scale also from the angular power spectrum.
Notice that the cosmological analysis in all of these studies was performed in harmonic space with the 
angular power spectrum, not in configuration space with the ACF full shape information.

In the present paper we focus on the less explored approach of using the full shape of the ACF  in configuration space 
to derive constraints 
on cosmological parameters, following the steps outlined in \citet{sobreira11}. 
We also estimate RSD parameters and define the BAO peak position using the method developed in \citet{Sanchez11}. 
For these purposes we  measure the ACF with the SDSS-III DR8 photometric data, using the so-called CMASS LGs catalogue \citep{RossEtAl}. 

This paper is organized as follows. 
In \S \ref{sec:data} we present the SDSS DR8 data to be analysed. 
In \S \ref{sec:measurecov} we briefly describe the novel method to estimate the ACF introduced by \citet{RossEtAl} and discuss how to construct the full covariance matrix including correlation among redshift shells. 
For completeness, in \S \ref{sec:acfmodel} we describe the theoretical modelling of the ACF. 
In \S \ref{sec:RSD} we find the best-fitting RSD parameters, and compare to the values found by \citet{crocceRSD} with a similar data set 
and also compare with BOSS DR9 spatial correlation function results \citep{Reid12}. 
The cosmological analysis is finally performed in \S \ref{sec:cosmo}, where we apply two methods. First,  we use the 
power law + Gaussian fit (PLG) approach first applied to real data by \citet{Carnero12}.
Secondly, for the first time using real data, we perform an estimation of cosmological parameters from the full shape 
information in the ACF. Finally, \S \ref{sec:conc} provides a summary and our conclusions. 

Throughout this study, when not stated otherwise, we assume as fiducial cosmological model a flat $\Lambda$cold dark matter ($\Lambda$CDM) universe with parameters as determined by WMAP7\footnote{During the final stages of this paper the WMAP9 results were released \citep{WMAP9}; as these results are similar to the previous WMAP7 ones we will continue to use the WMPA7 numbers since our findings would not be affected in a significant manner. Also, the Planck collaboration recently released its cosmological results \citep{Planckresults}; in this case it was found a significant difference with respect to WMAP7 mainly in $\Omega_m$ and the Hubble parameter. We comment on the impact of this difference on our results along the paper.
} \citep{wmap7}: dark matter density parameter $\Omega_{cdm} = 0.222$, baryon density parameter $\Omega_{b} = 0.0449$, Hubble parameter $h = 0.71$, 
primordial index of scalar perturbations $n_s = 0.963$, and normalization of perturbations $\sigma_8 = 0.801$. 
All numerical codes developed for our analysis applied the GSL package\footnote{http://www.gnu.org/software/gsl/}, and the linear matter power spectrum was computed with the {\tt CAMB} package \citep{CAMB}.

\section{The Data}\label{sec:data}

\subsection{Galaxy Selection}\label{sec:galsel}
In this work we use the imaging data from the SDSS DR8 \citep{Aihara2011}, which is publicly available by the SDSS team\footnote{http://portal.nersc.gov/project/boss/galaxy/photoz/}. This photometric sample has the same selection as the BOSS targets, which was intended to have approximately constant stellar mass, the so-called CMASS \citep{RossEtAl,White2011}. The construction of this photometric catalogue is detailed in \citet{RossEtAl} and \citet{Ho12}, where special care was taken to identify and remove potential systematic errors that could affect
the measurement of the angular clustering of galaxies. 

With the appropriate selection and cuts, one ends up with a catalogue containing $\sim 700$ thousand galaxies, mostly in the photometric redshift range $0.45\leq z_{p}\leq 0.65$, which is going to be our limiting redshifts for the cosmological analysis.  
Following \citet{RossEtAl} we will call this sample Luminous Galaxies, or LGs for short. 
We split the data in the range $0.45 \leq z_p \leq 0.65$ into 4 photo-z shells of width $\Delta z_p = 0.05$ and measure the  
ACF for each shell. We note that these are the same shells used in \citet{RossEtAl}, \citet{Ho12}, \citet{Seo12} and \citet{dePutter:2012sh}.

\subsection{Selection Functions}\label{sec:selfunc}

The true redshift distribution is one of the most important and challenging quantities needed in order to produce trustable 
results when investigating the projected angular clustering of galaxies within a redshift shell. 
In this sense, it is as important as the ACF measurement itself. 
For the LGs sample used in this work, the photo-zs of the objects are fairly accurate.  
They were estimated with the neural network ANNz code \citep{ANNz} using as training set 112,778 spectra, i.e. 
almost 10\% of the final photometric LGs sample.
The photo-z dispersion and the number of galaxies in each of the four shells are displayed in Table \ref{tb:photozNgal}. 

\begin{table}
\begin{center}
\begin{tabular}{ccc}
\hline
Redshift shell & $N_{gal}$ & $\sigma_{phot}$ \\
\hline
$0.45 \leq z_{p} \leq 0.50$ & 154531 &    $0.043$ \\
$0.50 \leq z_{p} \leq 0.55$ & 198132 &    $0.044$ \\
$0.55 \leq z_{p} \leq 0.60$ & 190603 &    $0.052$ \\
$0.60 \leq z_{p} \leq 0.65$ & 121181 &    $0.063$ \\
\hline
\end{tabular}
\caption{The four redshift shells used in this work. Columns show, for each shell, the photo-z range, the number of galaxies from \citet{Ho12} and the mean photo-z dispersion from \citet{RossEtAl}. }
\label{tb:photozNgal}
\end{center}
\end{table}

\begin{figure}
\begin{center}
  \begin{center}
    \includegraphics[width=3in]{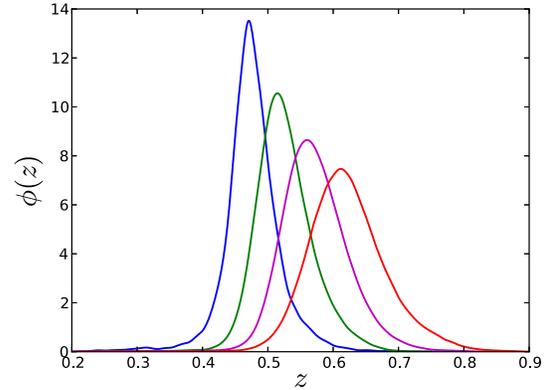}
  \end{center}
  \caption{Selection functions for the set of redshift shells applied in the cosmological analysis \citep{RossEtAl}.}
  \label{fig:selfunc}
\end{center}  
\end{figure}

The selection function convolves the redshift distribution with the photo-z errors and must be included in 
the ACF calculation as described in the next section.
In Fig.~\ref{fig:selfunc} we reproduce the selection functions $\phi(z)$ for the four redshift shells estimated by \citet{RossEtAl}, which is also publicly available. The selection functions overlap due to photo-z uncertainties, as expected. We properly account for this effect both in the ACF itself and in its covariance matrix, which accounts for the correlation amongst redshift shells, as explained in the next section. In order to speed up our numerical code to evaluate the theoretical ACF, we have smoothed the selection functions by applying a cubic spline, with an error below 0.01\%. 

\section{Measuring the ACF and Estimating its covariance}\label{sec:measurecov}
The estimation of the ACF was performed following \citet{RossEtAl}. 
We use the \textquoteleft $A_{star}$\textquoteright \, method (see section 4.1 of \citet{RossEtAl}) to 
correct for stellar density systematics and correct for the offset between SDSS photometry
in the North and South Galactic Cap \citep{Schlafly:2010dz} using the method applied to obtain their \textquoteleft $\Delta$ South\textquoteright \, results. Below we outline the main features for this evaluation.

The catalogue is pixelized at $N_{side} = 256$ using HEALPix \citep{healpix} and each pixel $i$ is assigned a weight $wt_i$ related to its
overlap with the imaging footprint.
The estimated ACF  $\hat{\omega}(\theta)$ is obtained from
\begin{equation}
  \hat{\omega}(\theta) = \frac{\sum_{ij}\delta_i \delta_j wt_i wt_j}{\sum_{ij}wt_i wt_j}\, ,
\end{equation}
where $\theta$ is the angular distance between pixel $i$ and pixel $j$ and 
the overdensity in pixel $i$, $\delta_i$, is given by 
\begin{equation}
  \delta_i = \frac{n_{i}}{\bar{n}wt_i} - 1\, ,
\end{equation}
where $n_i$ is the number of galaxies in pixel $i$ and $\bar{n} = \sum n_{i}/\sum wt_i$.

\begin{figure}
\begin{center}
\includegraphics[width=3in]{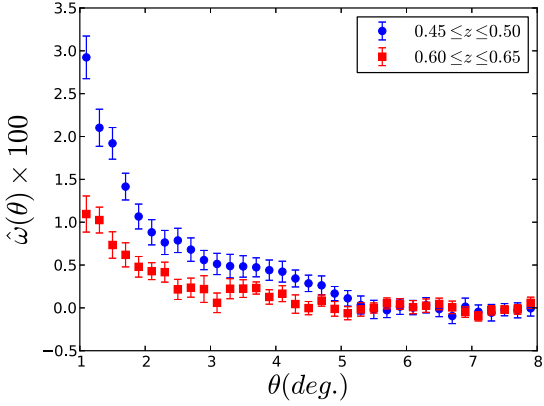}
\caption{Estimated ACF for the first and last redshift shells. The error bars are estimated via jackknife method.}
\label{fig:acf}
\end{center}
\end{figure}

In this work we measure the ACF in the angular range $1^\circ \leq \theta \leq 8^\circ$  with 35 angular bins for all redshift shells. 
This corresponds to spatial scales of $25 \lesssim r \lesssim 200\, h^{-1}$ Mpc. 
Note that in this approach, developed by \citet{RossEtAl} and \citet{Ho12}, it is straightforward to incorporate systematics effects, 
such as spurious clustering power due to extinction, seeing and star contaminations. 
In Fig. \ref{fig:acf} we show the ACF measurements for the first and last redshift shells. 
They show no excess of power at large scales found previously by  \citet{Thomas11b} and \citet{Sawangwit11} and the BAO peak is apparent in both shells.

It is well known that the estimation of the covariance matrix for a galaxy clustering analysis in configuration space is a difficult task. 
The standard way to construct the covariance matrix $C(\theta_i,\theta_j)$, between angular bins $i$ and $j$, is by the use of bootstrap methods, i.e., applying the data itself in the estimation. The most widely used approach is the so-called jack-knife method. 
The idea is to divide the survey into $N$ equal size areas, with the number of areas depending upon convergence tests, and producing the following errors (the covariance matrix is obtained with a similar procedure):
\begin{equation}\label{eq:jk}
  \sigma^{2}(\theta) = \frac{N-1}{N} \sum_{i=1}^{N} [\bar{\omega}(\theta) - \hat{\omega}_{i}(\theta)]^2 \, ,
\end{equation} 
where $\bar{\omega}(\theta)$ is the measured ACF for the full area and $\hat{\omega}_{i}(\theta)$ is the measurement when the $i$th jack-knife region is removed. The main shortcoming is that the jackknife method may give unstable results, especially at large scales. A noisy covariance matrix can change the best-fitting value for the parameters in a pronounced manner. Moreover, it does not give the covariance between redshift shells, which is needed when analysing the full sample of galaxies. We illustrate the importance of including the covariance in the next section, where we compare results obtained using the full covariance and using only diagonal errors.

In a recent study, \citet{CrocceEtAl} extensively studied a theoretical model for the ACF covariance matrix, assuming Gaussianity at large scale. They found that this approximation for the ACF covariance matrix at large scales is in very good agreement with the covariance from N-body simulations. This can be understood as a consequence of the central limit theorem, and of course, because at large scales one expects that the matter distribution follows a Gaussian distribution. Therefore, supported by this study, we will adopt the theoretical Gaussian covariance matrix in our analysis. As a bonus, for this covariance matrix it is straightforward to take into account correlation between redshift shells. 

The full Gaussian covariance matrix is given by (for a detailed description see, e.g.  \citealt{CrocceEtAl} and \citealt{sobreira11}):
\begin{eqnarray}\label{eq:fulcov}
 C^{\alpha,\beta}(\theta_i, \theta_j) &=& \frac{2}{f_{\mbox{\tiny sky}}} \sum_{l} \left[ \frac{2l+1}{(4 \pi)^2} P_{l}(\cos \theta_i) P_{l}(\cos \theta_j) \right.\\ \nonumber  
& & \left.\left( C_l^{\alpha,\beta} + 1/\bar{n}_{\alpha} \; \delta_{\alpha \beta} \right)^2 \right] 
\end{eqnarray}
The indices $\alpha$ and $\beta$ label the redshift shells. The angular power spectrum given by
\begin{equation}
C_l^{\alpha,\beta} = \frac{2}{\pi} \int dk\; k^2 P_{m}(k) \Psi_l^{\alpha}(k) \Psi_l^{\beta}(k).
\label{cl}
\end{equation}
In the above equations $f_{sky}$ is the fraction of the sky covered by the survey (in our case  $f_{sky} = 0.24$),
$P_{l}$ are Legendre polynomials, $\bar{n}_{\alpha}$ is mean density of galaxies in redshift shell $\alpha$,
$P_{m}(k)$ is the matter power spectrum, $\Psi_l^{\alpha}(k)$ is the kernel function due to RSD for redshift 
shell $\alpha$ and $\delta_{\alpha \beta}$ is the Kronecker delta, showing that the shot-noise enters only in the auto-covariance.

The cosmological parameters enter in the model for the covariance matrix through $P_{m}(k)$ and the kernel $\Psi_l^{\alpha}(k)$. In an ideal analysis one should construct the likelihood $\mathcal{L}$ with this information added in the best-fit search as was performed by \citet{Blake07}, i. e.,
\begin{equation}\label{eq:like}
  \mathcal{L} \propto |\mathbf{C}|^{-1/2}\, \mbox{exp}\left( -\frac{\mathbf{d}^{T}\mathbf{C}^{-1}\mathbf{d}}{2} \right)
\end{equation}
where $\mathbf{d} = \hat{\mathbf{\omega}} (\theta) - \mathbf{\omega}(\theta)$, is the vector with the difference between the measured ACF and 
its theoretical value for all redshift shells.  In our case, for four redshift shells we have $\hat{\mathbf{\omega}}(\theta) = (\hat{\omega}_1, \hat{\omega}_2,\hat{\omega}_3, \hat{\omega}_4)$, and $\mathbf{C}$ is the full covariance matrix with correlation between shells given in equation (\ref{eq:fulcov}). 
It is well known that the covariance matrix $\mathbf{C}$ is nearly singular, $|\mathbf{C}| \approx 0$, 
and we apply the singular value decomposition method \citep{numerical} in order to obtain its inverse. 
 
Another problem when applying the full likelihood method is that it is very time consuming to evaluate the theoretical covariance matrix for a given set of parameters, rendering a Markov chain Monte Carlo (MCMC) estimation of the parameters not viable. In order to overcome this issue we adopt the following strategy. We first fix a cosmology, in our case WMAP7, and assign an initial constant value $b=2$ for the bias, from which we generate a covariance matrix. 
Next we find the best-fitting value for the bias itself using the ACF full shape information as will be explained in the next section. In our case we find the following results for each redshift shell: $b = (1.94,\, 2.02,\, 2.15, \, 1.97)$. 

\begin{figure}
\begin{center}
\includegraphics[width=3in]{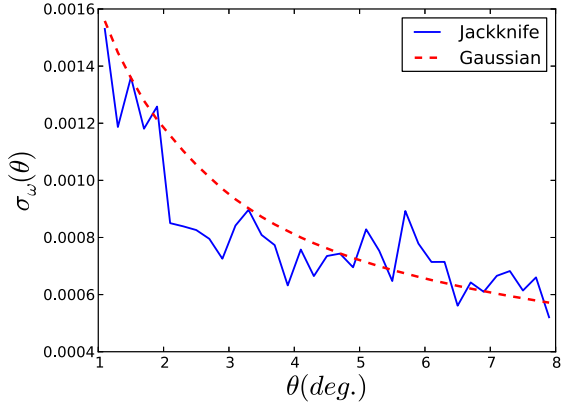}
\caption{Comparison between jackknife and theoretical errors for the redshift shell $0.55 \leq z_{p} \leq 0.60$.}
\label{fig:jk}
\end{center}
\end{figure}

With the fitted bias values, we compute the final covariance matrix that will be applied in our cosmological analysis. 
This approach assumes that most information in the covariance matrix comes from the bias or, in other words, most of the 
information in the covariance comes from the ACF amplitude and not its shape. 
In order to check the consistency of this assumption, we have compared the diagonal elements of the theoretical covariance matrix obtained 
using this procedure with the jack-knife results estimated with equation (\ref{eq:jk}) with $N=20$. 
In Fig. \ref{fig:jk} we show this comparison for the redshift shell $0.55 \leq z_{p} \leq 0.60$. 
The results are in fair agreement, giving some confidence on the use of the theoretical covariance matrix.
The consistency between the measured and theoretical 
covariance lead us to believe that our results would not change appreciably in a more complete analysis that 
vary the cosmology within the covariance matrix, or in a more conservative analysis that simply use 
the measured covariance.

Hence, we fix the cosmology when computing the covariance matrix, and we construct the standard $\chi^2$ statistics,
\begin{equation}
  -2\log\mathcal{L} = \chi^{2} = \mathbf{d}^{T}\mathbf{C}^{-1}\mathbf{d} \, 
\end{equation}   
to derive cosmological constraints from the data. This approach is widely applied in large scale clustering analysis, such as the analysis performed in the BOSS DR9 data, e.g. \citet{Bossresults1} and \citet{Bossresults2}.

\section{Modelling the angular correlation function}\label{sec:acfmodel}

Our modelling of the ACF is based on the methods used in previous studies (see, e.g. \citet{CrocceEtAl,ross11,sobreira11}). The ACF is related to the two-point spatial 
correlation function $\xi^{(s)}$ in redshift space by
\begin{equation}
\omega(\theta)=\int_0^\infty dz_1 \Phi(z_{1}) \int_0^\infty dz_2 \Phi(z_{2}) \xi^{(s)} \left( r(z_1,z_2,\theta)\right).
\label{angular}
\end{equation}
The function $\Phi(z)$ is determined by the selection function of the survey 
$\phi(z)$, the dark matter to luminous bias factor $b(z)$ 
and the linear growth function $D(z)$ (normalized to $D(z=0) =1$) 
as  $\Phi(z) = \phi(z) b(z) D(z)$.
The comoving distance $r(z_1,z_2,\theta)$ between two galaxies at 
redshifts $z_1$ and $z_2$ separated by an angle $\theta$ 
and in a flat cosmology is given by the relation, 
\begin{equation}
r = \sqrt{\chi^2(z_1)+\chi^2(z_2)-2\chi(z_1)\chi(z_2)\cos\theta},
\label{dist}
\end{equation}
where $\chi(z_i)$ is the radial comoving distance of the object $i$ to us (hereafter we use units with $c=1$):
\begin{equation}
\chi(z)=  \int_0^z\frac{dz'}{H(z')}
\end{equation}
and $H(z)$ is the usual Hubble function.

The redshift-space spatial correlation function $\xi^{(s)}$, in the plane-parallel approximation, is given by
\citep{Hamilton,Matsubara2000}
\begin{eqnarray}
\xi^{(s)}(r) &=&  \left[ 1 + \frac{1}{3} \left[ \beta(z_{1}) + \beta(z_{2}) \right] + \frac{1}{5} \beta(z_{1})\beta(z_{2}) \right] \xi_{0}(r)P_{0}(\mu) \nonumber \\ 
 &-& \left[ \frac{2}{3} \left[ \beta(z_{1}) + \beta(z_{2}) \right]
 + \frac{4}{7} \beta(z_1) \beta(z_2) \right] \xi_{2}(r)P_{2}(\mu) \nonumber \\ 
 &+& \left[  \frac{8}{35}\beta(z_{1})\beta(z_{2})\right] \xi_{4}(r)P_{4}(\mu) . 
\end{eqnarray}
Here the $P_{\ell}(\mu)$ are the usual Legendre polynomials as a function 
of $\mu=\hat{d} \cdot \hat{r}$ (cosine of angle between the line of sight $d$ and
$r$) and
$\beta(z) = f(z)/b(z)$ with $f(z) = d \ln D/d \ln a$.
The correlation multipoles are related to the matter power-spectrum $P_m(k)$ through:
\begin{equation}\label{xiP}
\xi_{l}(r) = \frac{1}{2\pi^{2}} \int_{0}^{\infty} dk k^{2} P_m(k) j_{l}(kr) \,\, ,
\end{equation}
and can be written as \citep{Hamilton}
\begin{equation}\label{monoham}
\xi_{0}(r) = \xi(r)\, ,
\end{equation}
\begin{equation}\label{quadham}
\xi_{2}(r) = \frac{3}{r^{3}}\int_{0}^{r}\!\! dx\, \xi(x)\, x^{2} - \xi(r)\, ,
\end{equation}
\begin{equation}\label{octoham}
\xi_{4}(r) = \xi(r) + \frac{5}{2} \left( \frac{3}{r^{3}}\int_{0}^{r} \!\! dx\, \xi(x)\, x^{2} \right) - \frac{7}{2} \left( \frac{5}{r^{5}}\int_{0}^{r} \!\! dx\, \xi(x)\, x^4\right) 
\, ,
\end{equation}
where $\xi(r)$ is the real-space spatial correlation function. One can incorporate the effects of non-linearities using the 
so-called renormalized perturbation theory (RPT) approach \citep{crocce08}, which determines the real-space correlation as
\begin{equation}
\xi_{nl}(r) = \xi(r) + A_{mc} \xi^{(1)}(r)\xi^{\prime}(r)\, ,   
\end{equation}
where $\xi^{\prime}$ is the derivative of $\xi(r)$ with respect to $r$ and
\begin{equation}
\xi^{(1)}(r) = \frac{1}{2\pi^{2}}\int_{0}^{\infty}dk k P_{m}(k) j_{1}(kr)  \, .
\end{equation}
For $A_{mc}$ we apply the value 1.55 found by \citet{CrocceEtAl} from N-body simulations. Another nonlinear effect that must be taken into account is the 
so-called Gaussian damping that affects mostly the BAO peak in the correlation function, and it is added  phenomenologically by introducing
a nonlinear power spectrum $P_{NL}(k)$ and substituting $P_{m}(k)$ as \citep{crocce08}:
\begin{equation}
P_m(k) \rightarrow P_{NL}(k) = P_m(k) \exp\left[-r_{NL}^2 k^2 D^2(z)/2 \right]
\end{equation}
with $r_{NL} = 6.6 $ Mpc h$^{-1}$. \citet{CrocceEtAl} showed that this approach is in good agreement with simulations on scales 
above $\sim 20 h^{-1}$Mpc; 
therefore our analysis will be applied above this scale.

In order to proceed it is worth pointing out some numerical issues that arise in going from $P(k)$ to $\xi(r)$. For this transformation, we have done some analysis varying the lower and upper limits in the integral (e.g. equation \ref{xiP}), since in principle it should be evaluated for all values of $k$. We found that with $k_{min} \simeq 10^{-5}\, h$Mpc$^{-1}$ and $k_{max} \simeq 100\, h$Mpc$^{-1}$, the integral converged and the time evaluation is reasonable for our purposes, something crucial for an extensive Markov chain analysis. Also, in order to compute the integrals in equations (\ref{quadham}) and (\ref{octoham}), we had to adopt a lower limit, and we found $r_{min} \simeq 0.01\,h^{-1}$Mpc to be a good value.

\section{Redshift-space distortion}\label{sec:RSD}

We start by using the LG data to examine the constraints on the parameters describing RSD following closely the study by \citet{crocceRSD}. 

In order to motivate the definition of the RSD parameters, we write the ACF in terms of a polynomial  in the bias $b$ and the velocity growth rate $f$ 
\begin{eqnarray}\label{eq:acfrsd}
  \omega(\theta) &=& b^{2}\omega_{0}(\theta) + bf \left( \frac{2}{3} \omega_{0}(\theta) + \frac{4}{3}\omega_{2}(\theta) \right) \nonumber \\
  &+& f^2 \left( \frac{1}{5}\omega_{0}(\theta) + \frac{4}{7}\omega_{2}(\theta) + \frac{8}{35}\omega_{4}(\theta) \right),
\end{eqnarray} 
where $\omega_{l}(\theta)$ is the projection of the space correlation function multipoles in the redshift shell. 
This equation is in fact an approximation of equation (\ref{angular}),  where one assumes that the functions 
$f(z)$, $D(z)$ and the bias do not evolve appreciably within each photo-z shell. We actually verified this assumption to hold in our case, by comparing both equations, (\ref{angular}) and (\ref{eq:acfrsd}), with our assumed cosmology and the LGs selection functions.

Since each term in equation (\ref{eq:acfrsd}) contains implicitly the product between $\sigma_8^{2}$ and $D(z)^{2}$, 
the two parameters that we are going to fit are
\begin{equation}
 b(z) \sigma_{8}(z) =  b\sigma_{8}D(z) 
\end{equation}
and
\begin{equation}
 f(z)\sigma_{8}(z) = f\sigma_{8} D(z) \, .
\end{equation}
In order to compare with \citet{crocceRSD}, in this section we adopt the following values for the cosmological parameters in a 
flat $\Lambda$CDM Universe: 
$\Omega_m = 0.272$, 
$\Omega_{b}=0.0456$, $n_s = 0.963$ and $h=0.704$. 
We slice the survey into four redshift shells between $z_p=0.45$ and $0.65$ with constant width $\Delta z_p = 0.05$, as defined in Section \ref{sec:galsel}.

We have constrained RSD parameters with three different approaches, all including correlations between shells due to photo-z dispersion. In the first approach we constrained $f\sigma_8$ and $b\sigma_8$ for each redshift shell, with a total of eight parameters. It should be noticed that within this approach the parameters best-fitting results are correlated. In the second approach we follow \citet{PaddyEtAl}, where it was noticed that the parameter $f\sigma_8$ does not change appreciably within the redshift limits in the $\Lambda$CDM model we adopt. Therefore, in this approach we fit only one $f\sigma_8$ parameter for all shells, but still allow $b\sigma_8$ to be different for each shell, for a total of five parameters to be fitted. In the third approach, we have assumed both $b\sigma_8$ and $f\sigma_8$ to be constant for all redshift shells, therefore we have only two free parameters. This last approach is similar to what is done with spectroscopic survey analysis, which typically find effective parameters over the whole survey range. The results for the three methods are respectively shown in Table \ref{tbl:rsd}.

\begin{table}
\begin{center}
\begin{tabular}{ccc}
\hline
Redshift shell & $b(z)\sigma_8(z)$ & $f(z)\sigma_8(z)$ \\
\hline
$0.45 \leq z_p \leq 0.50$ & $1.23 \pm 0.06$ & $0.66 \pm 0.33$ \\
$0.50 \leq z_p \leq 0.55$ & $1.25 \pm 0.11$ & $0.26 \pm 0.46$ \\
$0.55 \leq z_p \leq 0.60$ & $1.30 \pm 0.06$ & $0.93 \pm 0.37$ \\
$0.60 \leq z_p \leq 0.65$ & $1.16 \pm 0.08$ & $1.11 \pm 0.42$ \\
\hline
$0.45 \leq z_p \leq 0.50$ & $1.23 \pm 0.05$ & \multirow{4}{*}{$0.72 \pm 0.22$} \\
$0.50 \leq z_p \leq 0.55$ & $1.20 \pm 0.05$& \\
$0.55 \leq z_p \leq 0.60$ & $1.32 \pm 0.05$& \\ 
$0.60 \leq z_p \leq 0.65$ & $1.20 \pm 0.07$&\\
\hline
All shells & $1.24 \pm 0.04$ & $0.69 \pm 0.21$ \\
\hline
\end{tabular}
\caption{\label{tbl:rsd} Best-fitting values for RSD parameters for the three approaches: in the top part it is displayed the results allowing free parameters for each redshift shell; in the middle the velocity growth rate is assumed constant in all shells but allowing different $b(z)\sigma_8(z)$ for each shell and in the bottom part it is assumed constant $b(z)\sigma_8(z)$ and $f(z)\sigma_8(z)$ for all four shells.
}
\end{center}
\end{table}

The results for the first approach for the $b\sigma_8$ parameters are in good agreement with \citet{crocceRSD}, which reported $b\sigma_{8} = 1.26$, $1.21$ and $1.10$ with 2\% error for the first three shells (they did not analyse our last shell), although we find somewhat larger errors of $\sim 5\%$. If we assume $\sigma_8=0.8$, this translate to the following bias parameter for each shell: $b_1 = 1.94 \pm 0.08$, $b_2 = 2.02 \pm 0.08$, $b_3 = 2.15 \pm 0.08$ and $b_4 = 1.96 \pm 0.11$. Comparing with the results found by \citet{Ho12} we see that the first three shells agree quite well, only the last shell is 10\% lower. Although in principle the angular power spectrum $C_{l}$ and the ACF have the same information, in practice they can yield different results. This is mostly due to the need to include mask effects in the case of the angular spectra, and also in the estimation of the covariance matrix, which is nearly diagonal  for the power spectrum. The two analyses, performed independently, are complementary and the consistency between them 
provides an interesting cross-check of systematics. 

\begin{figure}
\begin{center}
\includegraphics[width=3in]{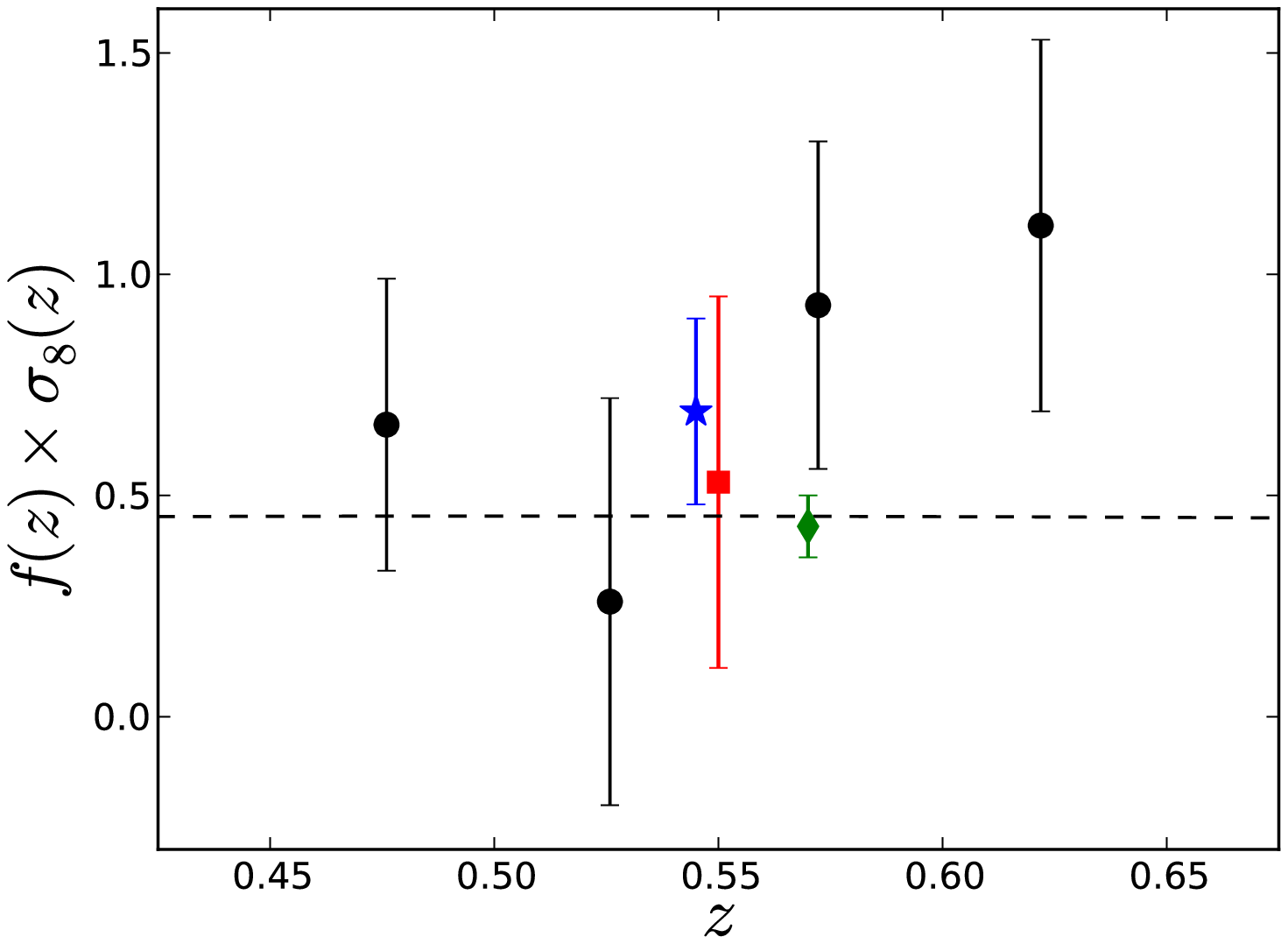}
\includegraphics[width=3in]{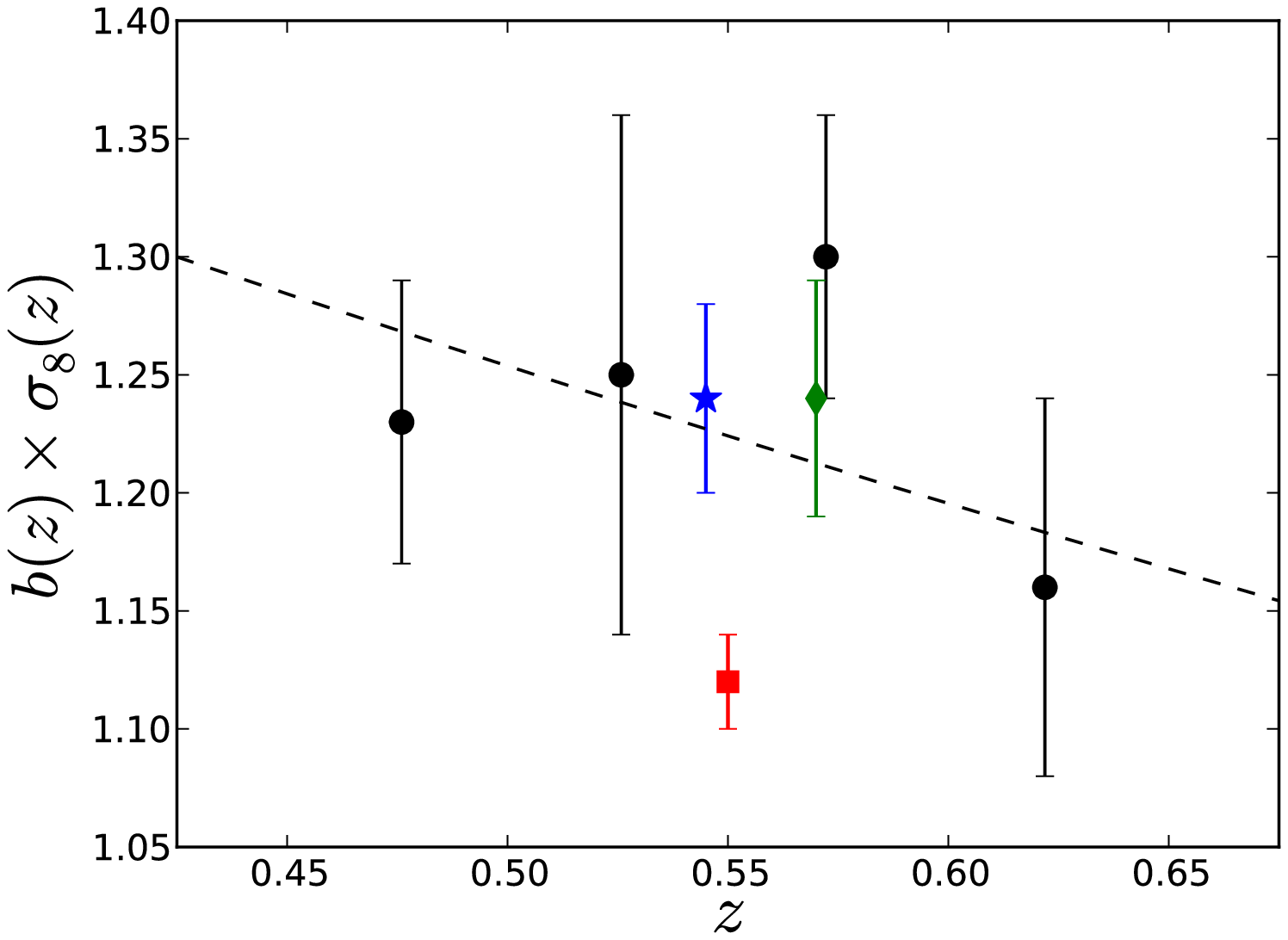}
\caption{\small Best-fit values for the RSD parameters for the first and third approaches as explained in the text: black bullets assuming different $f\sigma_8$ and $b\sigma_8$ for each shell and blue star assuming a constant $b\sigma_8$ and $f\sigma_8$ for all shells. The red squares are the results from \citet{crocceRSD} and green diamonds from \citet{Reid12}. The dashed lines are the expected theoretical values with $\sigma_8 = 0.8$ and $b=2$.}
\label{fig:rsd}
\end{center}
\end{figure}

For the product of the growth rate with $\sigma_8$, \citet{crocceRSD} found $f\sigma_8 = 1.14\pm 0.57$, $0.024\pm 0.53$ and $1.39\pm 0.46$ in redshift shells similar to our three lowest. Within the large errors, the results are compatible. 
In Fig. \ref{fig:rsd}, we show the best-fitting results for both parameters in each redshift shell. It shows that the first two shells are in agreement with the theoretical expectation within $1\sigma$ and the last two shells only agrees at the $2\sigma$ confidence level. For the bias parameters, it shows that only the third shell does not agree with a bias $b=2$ at $1\sigma$ level.

Recently, \citet{Reid12} reported the result for RSD parameters using BOSS DR9 spectroscopic data, which is a subset of the LGs we are analysing. They found $b\sigma_{8} = 1.24 \pm 0.05$ and $f\sigma_8 = 0.43 \pm 0.07$ at an effective redshift $z=0.57$ (their results are shown in Fig. \ref{fig:rsd} as a green diamond). The $b\sigma_{8}$ agrees quite well with our results for all approaches. When assuming a constant $f\sigma_8$ for all shells, we find $f\sigma_{8} = 0.72 \pm 0.22$ as our best-fitting value, which agrees with \citet{Reid12}  at $1.5\sigma$. For a better comparison with the \citet{Reid12} results, we analyse the results for our third method, in which case we find $b\sigma_{8} = 1.24 \pm 0.04$ and $f\sigma_8 = 0.69 \pm 0.21$ (both results are displayed in Fig. \ref{fig:rsd} as a blue star); in this case the bias is again consistent, but the velocity growth rate still agrees at the $1.5\sigma$ level. \citet{RossEtAl} showed that the highest redshift shell is most likely to be affected by systematic uncertainties, so we performed this last analysis without the last shell,  with results $b\sigma_{8} = 1.26 \pm 0.04$ and $f\sigma_8 = 0.64 \pm 0.23$, and in this case our results agree with \citet{Reid12} at $1\sigma$.

In a previous study, \citet{ross11} showed the impact of the assumed cosmology upon the RSD parameters using ACF. They found that changing $\Omega_m$ from $0.25$ to $0.30$ produces a significant effect on $f(z)\sigma_8(z)$ result. The difference between WMAP7 and Planck results \citep{Planckresults}, for a $\Lambda$CDM cosmology, is most pronounced in $\Omega_m$, which increases by $\sim 10\%$. Therefore the RSD is affected when assuming WMAP7 or Planck cosmologies. We re-analysed the data with Planck cosmology for the case of constant $f \sigma_8$ and $b \sigma_8$ in all four shells and found that $f\sigma_8$ increases by $33\%$ in comparison to WMAP7 cosmology, whereas $b \sigma_8$ does not change significantly, in agreement with \citet{ross11}.

In order to check the impact of the assumed cosmology on the covariance matrix we have also performed the fits with a different cosmology, varying $\Omega_m$, $\Omega_b$, $h$, within the WMAP7 allowed values. We found that the best-fitting results and the corresponding  $\chi^{2}$ do not change significantly, with a difference at the sub-percent level. Therefore we are confident that the approximation of keeping the theoretical covariance matrix fixed at a given cosmology does not bias our results significantly.

We also performed the analysis with only the diagonal errors, in angle and redshift, to check the impact of the covariance matrix on the results. The results for $f\sigma_{8}$ and its error are affected in a significant manner. The 
errors are typically four times smaller with respect to the errors with the full covariance matrix and the
$\chi^{2}$ are much higher for all shells. This demonstrates, as expected, that it is crucial to apply the full covariance matrix in the ACF analysis.

\section{Cosmological parameters}\label{sec:cosmo}

\subsection{PLG analysis}

In this section we apply the so-called PLG method \citep{Sanchez11} to extract the baryonic acoustic scale 
in the four redshift bins under analysis. The ACF is fitted around the BAO peak by a function of the form:
\begin{equation}
\omega(\theta)=A+B\theta^{\gamma}+C  e^{\frac{-(\theta - \theta_{plg})^2}{2 \sigma}}  \,\,  ,
\end{equation}
with six free parameters $(A,B,C,\gamma,\sigma, \theta_{plg})$.

We modify the PLG method by imposing some priors in the width of the BAO peak $\sigma$. The width of the BAO peak is defined by three factors: silk damping; adiabatic broadening of the acoustic oscillation and correlations of the initial perturbations. In configuration space, this correspond to $\approx 10\%$ of the BAO scale. Therefore we fix the width of the Gaussian to be proportional to $10\%$ of the BAO scale at a given redshift. We keep as free parameter the proportionality constant $p$ between the width and the mean of the peak, assumed to be independent of redshift. Therefore, from the original 24 free parameters, six for each shell, we have now five free parameters per redshift shell (20 in total, keeping $\sigma$ fixed), plus one extra free parameter $p$, related to the width of the peak by:
\begin{equation}
\sigma = 0.1 \, p \, \theta_{plg}  \,\,  .
\end{equation}

\begin{figure}
\begin{center}
\includegraphics[width=3in]{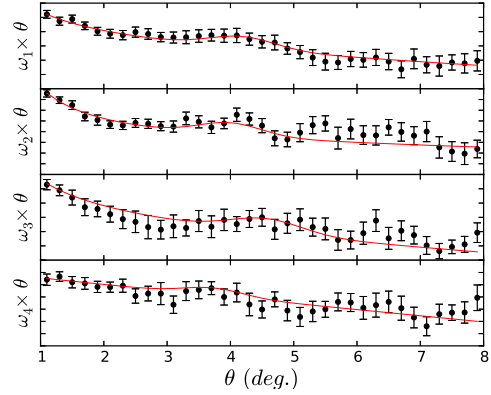}  
\caption{PLG fit (red line) using priors in the width of the Gaussian, to the four ACF (black dots) simultaneously using the full covariance matrix with correlations between redshift shells. 
We do not display the ACF values for clarity. The redshift shells are arranged from top to bottom with increasing redshift. We use this result as the best case.}
\label{fig:FIT}
\end{center}
\end{figure}

If we do not impose priors in $\sigma$, the PLG method can give unphysical results due to the noisy nature of the ACF measurement. For instance, in the second panel from top to bottom of Fig. \ref{fig:FIT}, a wide peak can be seen in the data around $\approx 3^{\circ}\!\!\! .6$, but with some decrease of the amplitude at the mean (consistent with noise), producing two peaks: one around $3^{\circ}\!\!\! .2$ and the other around $4^{\circ}\!\!\! .2$. The basic PLG method is not capable of solving this kind of structure, and for this reason we impose the prior discussed above.

The mean of the Gaussian $\theta_{plg}$ is associated with the true angular acoustic scale at redshift $z$, $\theta_{BAO}(z)$, through a correction $\lambda(z,\Delta z)$ that is independent of cosmology and only depends on redshift and redshift bin width:
\begin{equation}
\theta_{BAO}(z)=\lambda(z,\Delta z)\cdot \theta_{plg} \,\, .
\label{eq:correction}
\end{equation}
The parametrization of the function $\lambda(z,\Delta z)$ is described in \citet{Sanchez11}, where it is shown that this function does not vary significantly (sub-percent level) for 14 different cosmologies. 

In order to find $\Delta z$, one has to face a small difficulty. In a photometric analysis, one defines the top-hat bin width as the difference between the photo-z limits, which does not correspond to the true redshift bin width, due to the smearing by the photo-z error. Therefore, we need to correct the actual top-hat photo-z bin width to obtain the true bin width $\Delta z$. In our case, since we have the redshift selection function for each redshift bin, we can estimate the true redshift bin width for each redshift bin from the relation \citep{simpson09}
\begin{equation}
\Delta z=\sqrt{12}\, \sigma_z \,\, ,
\label{eq:truebin}
\end{equation}
where $\sigma_z$ is the dispersion of the selection function at the given redshift bin. In our analysis , we obtain a true bin width for each bin given by $\Delta z_{true}^{1} = 0.101$, $\Delta z_{true}^{2} = 0.130$, $\Delta z_{true}^{3} = 0.158$ and $\Delta z_{true}^{4} = 0.185$. 

These measurements can then be used to constrain the angular diameter distance as a function of redshift, related by:
\begin{equation}
\theta_{BAO}(z)=\frac{r_{s}}{D_{A}(z)} \,\, ,
\end{equation}
 extracted from the ACF, where $r_{s}$ is the baryonic acoustic scale at decoupling and $D_{A}(z)$ is the angular diameter distance at redshift $z$.

The four redshift bins are fitted together considering the covariance matrix between redshift shells introduced in section \ref{sec:measurecov} using the \textit{Minuit} library \citep{minuit}. In total, there are 21 free parameters, with a number of 140 data points and hence 119 degrees of freedom. In Fig. \ref{fig:FIT}, the fit results are shown on top of the measured ACF for the four redshift bins and in Table \ref{tab:FIT} we display the main values of the fit. The best-fitting value for the proportionality constant is $p=1.28$, resulting that the width of the acoustic peak is $12.8\%$ the scale of the peak, common to all redshift bins and in agreement with what is expected from theory.

\begin{table}
\begin{center}
\begin{tabular}{cccc}
\hline
Redshift shell & $\theta_{plg}\,(^{\circ})$ &$\theta_{BAO}\,(^{\circ})$ & $S/N$ \\
\hline
$0.45 \leq z_{p} \leq 0.50$ &  $4.13 \pm 0.19$ & $4.48 \pm 0.29$ & 2.6\\
$0.50 \leq z_{p} \leq 0.55$ &  $3.93 \pm 0.20$ & $4.28 \pm 0.29$ & 2.5\\
$0.55 \leq z_{p} \leq 0.60$ &  $4.49 \pm 0.31$ & $4.90 \pm 0.39$ & 2.0\\
$0.60 \leq z_{p} \leq 0.65$ &  $3.68 \pm 0.23$ & $4.01 \pm 0.30$ & 1.3\\
\hline
\end{tabular} 
\caption{PLG fit results for the four redshift bins imposing priors in the width of the BAO peak. The overall quality of the fit is $\chi^{2}/dof = 0.96$. $\theta_{BAO}$ is obtained after correcting from projection effects (equation \ref{eq:correction}), and its error accounts for both statistical and systematic errors. 
The signal-to-noise ratio (S/N) is giving as the strength of the Gaussian divided by its error (in our parametrization, parameter $C$).}
\label{tab:FIT}
\end{center}
\end{table}

After correcting from projection effects (using the true redshift bin width), we obtain $\theta_{BAO}$ for each redshift bin, as shown in Table \ref{tab:FIT}. Errors in $\theta_{BAO}(z)$ have two main contributions: the statistical error coming from the fit plus an intrinsic error due to the variance introduced by the photometric redshift uncertainty, estimated to be around 5\% for a SDSS like survey \citep[details are found in][]{Carnero12}. The combined errors are presented in Table \ref{tab:FIT}.

\begin{figure}
\begin{center}
\includegraphics[width=3in]{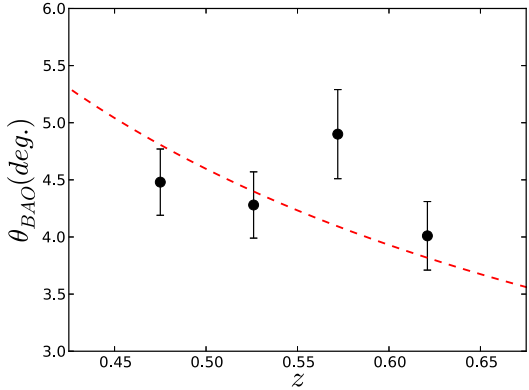}
\caption{$\theta_{BAO}$ as a function of $z$ for the CMASS catalogue. The dashed line is given by the best-fitting cosmology, when the $\theta_{BAO}$ measured in the CMB is also used. The best-fitting cosmology is $\Omega_{m}=0.249 \pm 0.031$, $w=-0.885 \pm 0.145$, fixing the other parameters to $h=0.71$ and $\Omega_{b}=0.0449$.
}
\label{fig:bao}
\end{center}
\end{figure}

In Fig. \ref{fig:bao} the evolution of $\theta_{BAO}$ as a function of redshift is shown for our analysis, together with the best-fitting value stated below, when the $\theta_{BAO}$ measured in the cosmic microwave background (CMB) is also used. Errors are given as the diagonal term in the full covariance matrix of the parameters for the four redshift shells (statistical plus systematic error).

To find the best-fitting cosmology we parametrize $r_{s}$ as a function of cosmological parameters using the analytical approximation given in \citet{eisenstein97}. We then minimize the $\chi^2$ statistics using the four BAO measurements together with the BAO measurement at decoupling measured by WMAP7, with $\theta_{BAO}(z=1091)=0.5952 \pm 0.0016$ degrees. The best-fitting cosmology for free parameters $\Omega_{m}$ and $w$, fixing the other parameters to $h=0.71$ and $\Omega_{b}=0.0449$ is $\Omega_{m}=0.249 \pm 0.031$, $w=-0.885 \pm 0.145$. If we instead use only the DR8 measurements in the ($\Omega_{m}$, $w$) space, there are not enough degrees of freedom and the cosmology is poorly constrained. Therefore, we fit only $\Omega_{m}$ for a $\Lambda$CDM model, with a best-fitting result of $\Omega_{m}=0.231 \pm 0.079$.

Throughout the analysis we have fixed the effective number of neutrino species to $N_{eff}=4.34$, the central value of the result found by WMAP7 data in combination with BAO and $H_{0}$ priors \citep{wmap7}, which deviates from what is expected in the standard model of particle physics ($N_{eff} = 3.04$). In order to test the effect of $N_{eff}$ on the cosmological constraints, we find the best-fitting value to $\Omega_{m}$ using the BAO measurements from DR8 alone, with $N_{eff}=3.26$, obtained in WMAP9 \citep{WMAP9}. In this case, the best fit is $\Omega_{m}=0.292 \pm 0.090$, $\sim 25\%$ higher than with WMAP7 $N_{eff}$. This result shows that the effective number of neutrino species is an important parameter in the analysis of the BAO if we use it as a standard ruler, and its uncertainty will need to be considered in future analysis.

In order to compare our results with previous measurements, we use the results from \citet{Seo12}, where the BAO position for the same data was obtained  from the angular power spectrum with a different methodology and stacked the $C_{l}$ for each shell together to give a single angular distance measurement at $z=0.54$. They measured the deviation of the best-fitting cosmology with reference to a fiducial model with $\Omega_{m}=0.274$, $w=-1$, $\Omega_{b}=0.049$, $h=0.7$, parametrized by: 
\begin{equation}
\alpha=D_{A}(z)/{D_{A}(z)}_{fiducial}
\end{equation}
The results found by \citet{Seo12} was $\alpha=1.066 \pm 0.047$, after marginalizing over the other cosmological parameters. Our parametrization is different, and we did not calculate a stacked ACF from the four redshift bins. Nonetheless, we can also measure the deviation from the same fiducial cosmology and obtain a value for $\alpha$ using the values obtained with the PLG method. In this case we do not use the BAO measurement from WMAP7. Without marginalizing, and using the best-fitting cosmology, we obtain a value of $\alpha=1.028 \pm 0.035$. This value is roughly at $1\sigma$ from the value of \citet{Seo12}.  The error is smaller in our case because we have not marginalized over the other parameters. We note that the DR9 BOSS study \citep{Anderson:2013oza}
found that $1\sigma$ differences between the BAO position recovered from the power spectrum
and that recovered from the correlation function were not unusual.

\subsection{Full shape analysis}
\begin{table}
\begin{center}
\begin{tabular}{cccc}
\hline
Redshift shell & $\Omega_m$ & $f_{b}$ & $\chi^{2}/\mbox{dof}$ \\
\hline
$0.45 \leq z_{p} \leq 0.50$ & $0.29 \pm 0.04$ & $0.25 \pm 0.04$ & 0.65\\
$0.50 \leq z_{p} \leq 0.55$ & $0.37 \pm 0.05$ & $0.14 \pm 0.04$ & 1.49\\
$0.55 \leq z_{p} \leq 0.60$ & $0.25 \pm 0.04$ & $0.13 \pm 0.04$ & 0.79\\
$0.60 \leq z_{p} \leq 0.65$ & $0.23 \pm 0.04$ & $0.23 \pm 0.06$ & 0.62\\
\hline
All shells combined & $0.280 \pm 0.022$ & $0.211 \pm 0.027$ & 1.05 \\
\hline
\end{tabular}
\caption{\label{tbl:cosmofit} Best-fitting values from the full-shape ACF analysis for $\Omega_{m}$ and $f_{b} = \Omega_{b}/\Omega_{m}$, after marginalizing over $\sigma_8$ and bias, with all other parameters being kept fixed at the WMAP7 cosmology as stated in the text. }
\end{center}
\end{table}
In this section we apply the full shape information of the ACF to constrain a subset of cosmological parameters, namely  
$\Omega_m$, $f_b = \Omega_b/\Omega_m$, $\sigma_8$ and bias. Following \citet{Blake07} and \citet{Thomas11a}, we assume the bias to be scale independent and constant within each shell.
The other cosmological parameters are held fixed at the WMAP7 values given previously. 
The $\chi^2$ function was constructed as discussed in section \ref{sec:measurecov} and we used the 
{\tt CosmoMC} package \citep{cosmomc} to search the parameter space, with no priors imposed in any of the free parameters.

First we consider the behaviour of the parameters $\sigma_8$ and bias $b$. These two parameters are highly degenerate, making it difficult to constrain them separately \citep{okumura}. Therefore we show the constraint of their product for each redshift combined: $\sigma_8 b_1 = 1.46 \pm 0.09$, $\sigma_8 b_2 = 1.49 \pm 0.10$, $\sigma_8 b_3 = 1.63 \pm 0.12$ and $\sigma_8 b_4 = 1.51 \pm 0.15$. In order to compare with the previous results from the RSD analysis (see section \ref{sec:RSD}) we assume $\sigma_8 = 0.801$ as before: $b_1 = 1.82 \pm 0.12$, $b_2 = 1.86 \pm 0.13$, $b_3 = 2.03 \pm 0.15$ and $b_4 = 1.89 \pm 0.17$. Although the results from the full shape analysis are typically $5\%$ lower, they all agree at the $1\sigma$ level.

\begin{figure*}
\begin{center}
\includegraphics[width=6.6in]{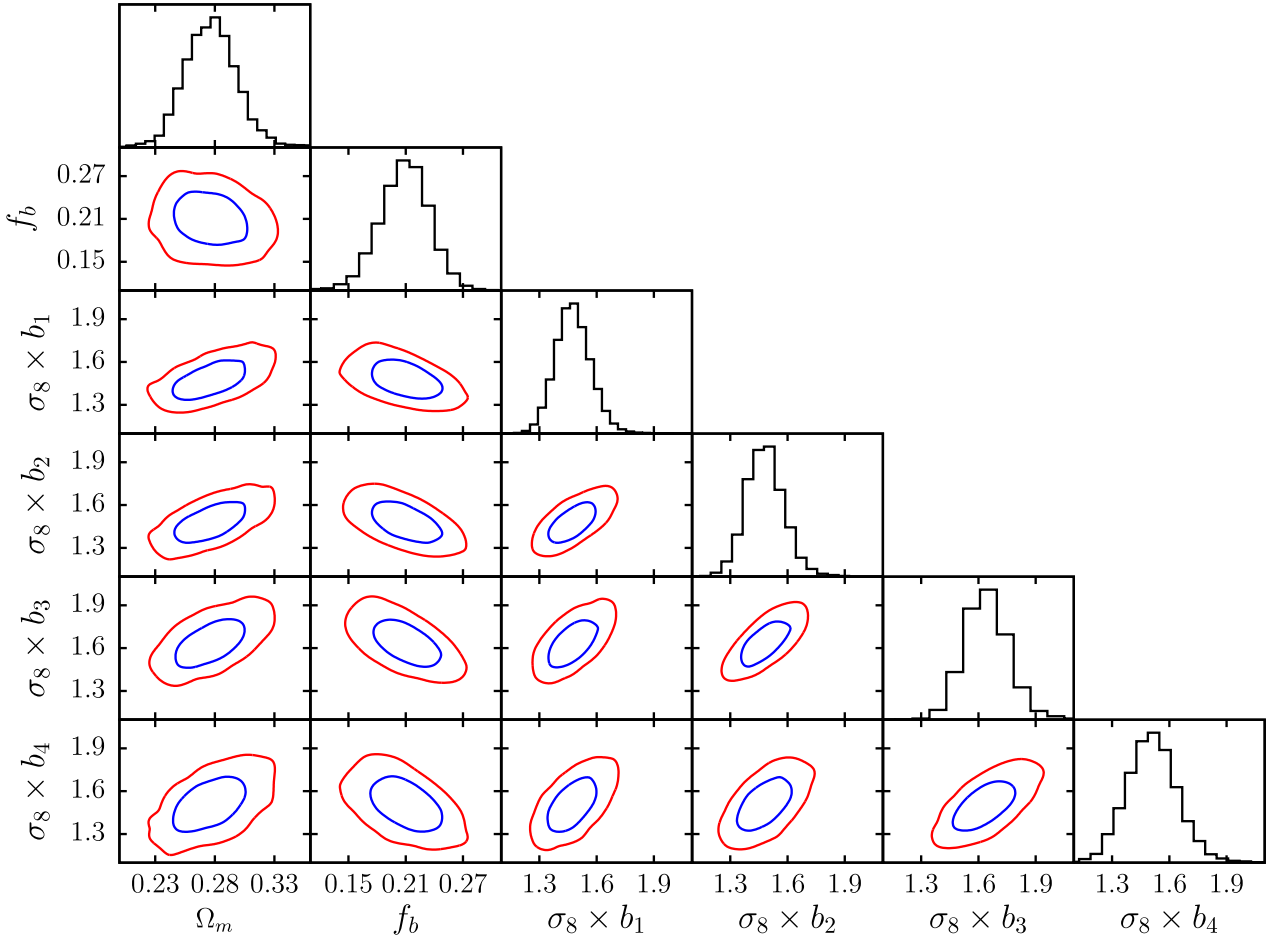}
\caption{Likelihood contours for $(\Omega_m$, $f_b$, $\sigma_8 b_1$, $\sigma_8 b_2$, $\sigma_8 b_3$, $\sigma_8 b_4)$ with all other parameters being kept fixed in WMAP7 cosmology. The diagonal panels display the marginalized likelihood for each one of the six parameters. The other plots show the $1\sigma$ and $2\sigma$ confidence regions of each pair of parameters, with the other marginalized.} 
\label{fig:cosmofit}
\end{center}
\end{figure*}

We now focus our attention on other parameters in our cosmological analysis, namely $\Omega_m$ and $f_b$. First we perform the analysis for each shell independently in order to check the dispersion of the best-fitting values. In this case we have four free parameters. In Table \ref{tbl:cosmofit}  we show our results. Both parameters vary appreciably among different redshift shells, but they all agree within $2\sigma$. Even though the dispersion is non-negligible, the best-fit values oscillate around the expected results coming from WMAP7 namely, $\Omega_m = 0.266$ and $f_b = 0.17$. 
This result already indicates that the combination of all shells will give results in agreement with WMAP7, anticipating the main results of this analysis. When comparing with the analysis from \citet{Blake07} and \citet{Thomas11a}, which used the angular power spectrum, we found that our results are compatible in all shells. Compared to the results of \citet{Thomas11a}, the errors on the parameters that we find are smaller, because the area and number of galaxies in our data set are larger. 

\begin{figure}
\begin{center}
\includegraphics[width=3in]{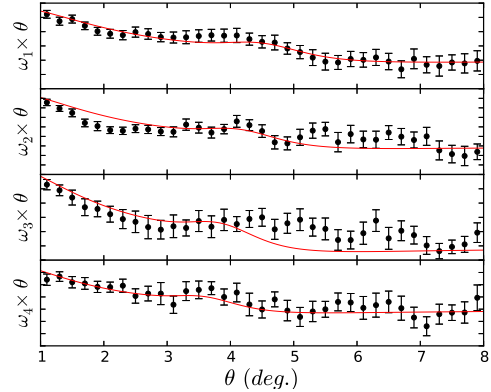}  
\caption{Best-fitting ACF full shape information when combining all shells given in the last line of Table \ref{tbl:cosmofit} (red line) together with the measurements (black dots). Not only the angular bins are correlated but also the ACF between shell. We do not display the ACF values for clarity. The redshift shells are arranged from top to bottom with increasing redshift.}
\label{fig:acffull}
\end{center}
\end{figure}

The analysis for all combined redshift shells also accounts for the correlation among shells. For this analysis we have seven free parameters, the best-fitting results for $\Omega_m$ and $f_b$ are displayed in Table \ref{tbl:cosmofit} and the marginalised probability distribution function and the 2D likelihood contours are shown in Fig. \ref{fig:cosmofit} for all parameters analysed. The results are: $\Omega_m = 0.280 \pm 0.022$ and $f_b = 0.211 \pm 0.026$ which translates into $\Omega_b = 0.059 \pm 0.008$. The matter density parameter found in our analysis is in good agreement with the value from WMAP7, with a difference of $5\%$. The baryon fraction best-fit is higher than WMAP7, $\sim 20\%$, but in agreement within $1\sigma$. In Fig. \ref{fig:acffull} we show the ACF measurements together with the best-fitting ACF for this analysis. It shows that the model is in good agreement with the measurements and the BAO peak is evident for three redshift shells as already shown in the previous section. The higher $\chi^2$ for the second shell is evident due to the poor fit at small scales.   

As stated in the previous section, the main difference between WMAP7 and Planck results is $\Omega_m$ and the Hubble parameter. The former is $\sim 10\%$ higher with Planck's data and the latter is $\sim 4\%$ lower. As $\Omega_m$ is left as a free parameter in our analysis, it is not an issue, but $h$ is fixed. As shown in \citet{Blake07} the major effect of changing $h$ is in the $\Omega_m$ best fit. Because the clustering characteristics are driven mostly by the combination $\Omega_m h$,  lowering $h$ implies in the increase of $\Omega_m$. Since our best-fit with WMAP7 Hubble parameter is $\Omega_m = 0.280$, if we instead use $h = 0.68$, as found by Planck, we would have found a higher value, in better agreement with $\Omega_m$  quoted by Planck.

As an additional cross-check of the results in this section, we have repeated the analysis above using a completely 
independent set of codes, both for the estimation of the theoretical ACF as well as for the Monte Carlo sampling. 
We coupled the independent ACF code to the {\tt emcee} sampler \citep{emcee} and repeated all the calculations. We find that the results obtained from {\tt CosmoMC} and {\tt emcee} are in most cases nearly identical and in all cases consistent with each other 
within 1$\sigma$ errors. 

The quoted errors of $8\%$ for $\Omega_m$ and $12\%$ for $f_b$ are underestimated since they do not take into account 
the marginalization of the other parameters. For more realistic errors, we should have varied all parameters, including the Hubble parameter $h$, spectral index $n_s$ and the dark energy equation of state parameter $w$, and marginalized over them. Unfortunately the statistical significance of our data set alone is not sufficient to obtain useful constraints. Combining our results with a CMB likelihood, e.g. from WMAP or Planck, would probably allow for a 
more complete analysis and for better constraints due to the complementarity of these probes (see e.g. \citet{Ho12}.

Nonetheless, our results point out that the methods applied to extract information 
from measurements of ACF in configuration space are able to yield competitive cosmological constraints. 
This indicates that these methods will be even more useful when applied to future data sets with greater constraining power.
The combination with other probes of large-scale structure and CMB should provide additional 
consistency checks and even better constraints.  

\section{Summary and Conclusions}\label{sec:conc}

We have analysed the large-scale ACF of luminous galaxies from the SDSS-DR8 photometric data. 
The ACF was measured in four photo-z shells with the novel approach develop by \citet{ross11} and \citet{Ho12}, which incorporates 
systematics effects and was able to remove the excess of power at large scales as reported by previous studies \citep{Sawangwit11,Thomas11b}.

We have performed three different analyses using the measured ACFs: RSD; BAO detection using the PLG method and a cosmological analysis with the ACF full shape information. The latter represents, to the best of our knowledge, the first cosmological analysis performed with the ACF in configuration space. All three analyses accounted for the correlation between redshift shells and effects of photo-z errors encoded in the selection function. 
Our main results are the following.

\begin{itemize}
  \item{Within the redshift space parameters best-fit and assuming $\sigma_8 = 0.801$, we found that the bias parameters are in good agreement with other DR8 measurements, such as those by \citet{Ho12}.}
  \item{When allowing for arbitrary values of $f\sigma_{8}$ in each redshift shell, the RSD parameters vary appreciably around the expected value from $\Lambda$CDM cosmology, in agreement with the findings of \citet{crocceRSD}.}
  \item{When assuming constant RSD parameters over the survey range we found $b\sigma_8 = 1.24 \pm 0.04$ and $f\sigma_8 = 0.69 \pm 0.21$. The bias parameter agrees quite well with BOSS DR9 measurements \citep{Reid12}, and the growth rate agrees within $1.5\sigma$.}
  \item{We extracted the position of the BAO peak using the PLG parametrization for all four shells, and combined these measurements 
  with the BAO peak in the CMB data from WMAP7. 
  We obtained cosmological constraints of $\Omega_{m}=0.249 \pm 0.031$ and $w=-0.885 \pm 0.145$. 
  For a $\Lambda$CDM model, and using only our own ACF measurements, we obtained $\Omega_{m}=0.231 \pm 0.079$ with other parameters fixed at our fiducial cosmology.}
    \item{Within the ACF full shape analysis we constrained $\Omega_{m}$ and $f_b$ for each redshift shell independently, and found that the best-fit values oscillate around the WMAP7 values, but are all within $2\sigma$.}
  \item{When analysing all shells combined, with the full covariance matrix accounting for the redshift correlations, 
  the best-fit values were: $\Omega_{m} = 0.280 \pm 0.022$ and $f_b = 0.211 \pm 0.026$ in reasonable agreement with WMAP7.}
  \item{Both analysis performed in this work to constrain cosmology, namely the BAO peak and full shape information, agree in the 
  $\Omega_m$ best-fit values, showing that both methods are consistent with each other.}
\end{itemize}

We have shown that the ACF estimated from photometric data can be efficiently applied to constrain cosmological parameters. The ACF results for the photometric DR8 data are clearly not as competitive as those 
from the spatial correlation function, which already provides stronger constraints with the BOSS DR9 data \citep{Bossresults2,Bossresults1}. Nonetheless,  our results are encouraging for future photometric surveys, such as the DES, PanSTARRS and LSST, which will probe larger redshifts and measure significantly more galaxies.  In this case, the ACF measurements have the potential to accurately constrain a  larger number of cosmological parameters \citep{sobreira11}, allowing for extra consistency checks with other independent probes.

\section*{Acknowledgements}

We thank the anonymous referee for useful comments on the manuscript. We thank Joe Zuntz for useful discussions on MCMC codes and Shirley Ho for comments on the draft paper.
FdS would like to thank Mariana Penna-Lima and Sandro Vitenti for valuable interactions.

AC and FS acknowledge financial support from CNPq (PCI-D grants 311876/2011-0 and 303186/2011-9 respectively, 
both associated with the PCI/MCT/ON Program).
HOC is supported by CNPq.
ML and RR are partially supported by FAPESP and CNPq.

This research was carried out with the support of the Laborat\'orio Interinstitucional de e-Astronomia (LIneA)
operated jointly by the Centro Brasileiro de Pesquisas F\'isicas (CBPF), the Laborat\'orio Nacional de Computa\c c\~ao Cient\'ifica (LNCC) and 
the Observat\'orio Nacional (ON) and funded by the Minist\'erio de Ci\^encia e Tecnologia (MCT).

Funding for SDSS-III has been provided by the Alfred P. Sloan Foundation, the Participating Institutions, the National Science Foundation, and the US Department of Energy Office of Science. The SDSS-III web site is http://www.sdss3.org/.

SDSS-III is managed by the Astrophysical Research Consortium for the Participating Institutions of the SDSS-III Collaboration including the University of Arizona, the Brazilian Participation Group, Brookhaven National Laboratory, University of Cambridge, Carnegie Mellon University, University of Florida, the French Participation Group, the German Participation Group, the Instituto de Astrof\'isica de Canarias, the Michigan State/Notre Dame/JINA Participation Group, Johns Hopkins University, Lawrence Berkeley National Laboratory, Max Planck Institute for Astrophysics, New Mexico State University, New York University, Ohio State University, Pennsylvania State University, University of Portsmouth, Princeton University, the Spanish Participation Group, University of Tokyo, University of Utah, Vanderbilt University, University of Virginia, University of Washington, and Yale University.

\bibliographystyle{mn2e}

\label{lastpage}
\end{document}